\documentclass[a4paper,11pt]{article}
\pdfoutput=1
\usepackage{color}
\definecolor{MyDarkBlue}{rgb}{0,0.08,0.45}
\usepackage{hyperref}
\hypersetup{
colorlinks=true,
citecolor=MyDarkBlue,
linkcolor=MyDarkBlue,
urlcolor=MyDarkBlue,
pdfauthor={Kevin Goldstein and Hossein Yavartanoo},
pdftitle={A  note on non-linear electrodynamics, regular black holes and the entropy
  function},
pdfsubject={hep-th,gr-qc}
}

\usepackage{amsmath,amssymb}
\usepackage{graphicx}
\numberwithin{equation}{section}
\usepackage[numbers,sort&compress]{natbib}
\usepackage{hypernat}
\newcommand{\be}{\begin{equation}}
\newcommand{\bea}{\begin{eqnarray}}
\newcommand{\eea}{\end{eqnarray}}
\newcommand{\ba}{\begin{array}}
\newcommand{\ea}{\end{array}}
\newcommand{\ee}{\end{equation}}
\newcommand{\p}{\partial}
\newcommand{\refeq}[1]{\stackrel{(\ref{#1})}{=}}
\newcommand{\F}{\mathcal{F}}
\newcommand{\G}{\mathcal{G}}

\usepackage[cm]{fullpage}
\usepackage{ifpdf}
\ifpdf
\DeclareGraphicsExtensions{.pdf}
\else
\DeclareGraphicsExtensions{.eps}
\fi

\begin{document}

\begin{flushright} \small
SPIN-07/47\\ ITP-UU-07/61
\end{flushright}
\bigskip

\begin{center}
 {\LARGE\bfseries A  note on non-linear electrodynamics, regular black holes and the entropy function.}  
\\[10mm]
\textbf{ Kevin
  Goldstein\footnote[1]{\tt k.goldstein [at] uu.nl}
  and Hossein
  Yavartanoo\footnote[2]{\tt yavar [at] phya.snu.ac.kr}
}\\[5mm]
{$^{1}$\em Institute for Theoretical Physics,\\ Utrecht University, Utrecht, The Netherlands}\\[3mm] 
$^{2}$ {\em Center for Theoretical Physics and BK-21 Frontier Physics Division, \\
 Seoul National University, Seoul 151-747 KOREA} 
\end{center}

\vspace{5mm}

\bigskip
\centerline{\bfseries Abstract} 
\medskip
\noindent
We examine four dimensional magnetically charged extremal black holes in certain
non-linear $U(1)$ gauge theories coupled to two derivative gravity. For a given coupling,
one can tune the magnetic charge (or vice versa) so that the curvature singularity at the
centre of the space-time is cancelled. Since these solutions have a horizon but no
singularity, they have been called regular black holes.  Contrary to recent claims in the
literature, we find that the entropy function formalism reproduces the near horizon
geometry and gives the correct entropy for these objects. 
\bigskip

\setcounter{footnote}{0}

\section{Introduction}
\label{intro}

The Penrose cosmic censorship hypothesis states that, if singularities predicted by
General Relativity occur in nature, they must be dressed by event horizons
\cite{Hawking:1973uf}.  Behind the veil of an event horizon, there is no causal contact
from the interior to the exterior of a black hole, so the pathologies occurring at the
singular region can have no influence on an external observer. However, the converse of
the hypothesis is apparently not true --- a horizon does not necessarily hide a
singularity. Solutions with a horizon but no singularity have been called regular black
holes.
  
The holographic principle, \cite{gr-qc/9310026,9409089}, states that the number of
degrees of freedom describing the black hole is bounded by the area of the horizon. A
stronger statement is that degrees of freedom living on the horizon can describe the
physics of the interior completely. While the holographic principle is essentially a
proposed feature of quantum gravity, one might wonder whether having a classically regular
or singular solution has any quantitative or qualitative effect on the entropy and the
physics at the horizon.

Earlier work on regular black hole models can be found in
\cite{bardeen1968pg,Barrabes:1995nk,gr-qc/9403049,Mars:1996np,Cabo:1997rm} These regular
solutions are referred to as “Bardeen black holes” \cite{Borde:1996df}. In addition,
regular black hole solutions to Einstein equations with various physical sources were
reported in \cite{AyonBeato:1998ub} and \cite{Magli:1997mw}. Among known regular black
hole solutions,  are the solutions to the coupled equations of
nonlinear electrodynamics and general relativity found by Ay\'on-Beato and Garci\'a
\cite{hep-th/9911174} and by Bronnikov \cite{gr-qc/0006014}. The latter describes
magnetically charged black hole, and provides an interesting example of the system that
could be both regular and extremal.  In this note we are specially interested in the near
horizon geometry of an extremal magnetically charged black hole non-linearly coupled to a
$U(1)$ gauge field. For a given magnetic charge, one can tune the non-linear coupling so
that the solution is regular. 

In \cite{gr-qc/0403109}, Matyjasek found the near horizon, $AdS_2\times S^2$ geometry of a
particular magnetically charged extremal black hole.  The entropy function formalism of
Sen \cite{hep-th/0506177,hep-th/0508042}, is particularity useful for discussing the
entropy of extremal black holes especially when non-linear or high derivative terms make a
full analysis difficult. Since, the formalism is equivalent to solving Einsteins equations
for the near horizon region, a priori, and assuming the near horizon geometry decouples,
one would expect to be able to reproduce the results of \cite{gr-qc/0403109} using Sen's
approach.  This issue has been studied  recently,  \cite{0705.2478,0707.1933}, and
authors reported that, even at the level of two derivative gravity, the entropy function
approach does not lead to the correct Bekenstein-Hawking. To account for this discrepancy,
they claim that the entropy function approach is sensitive to whether the nature of the
central region of the black hole is regular (linear) or singular (nonlinear).

Contrary to the claims of, \cite{0705.2478,0707.1933}, in this note we find that a straight
forward application of the entropy function formalism reproduces the results of
\cite{gr-qc/0403109}.  The equation of motion derived from extremizing the entropy
function are exactly the same as equation of motion at horizon found by extremizing
the action, since the entropy function (up to Legendre transformation) is the Lagrangian
at the horizon. The fact that the entropy is the value of entropy function at its extremum
is derived from the Wald entropy formula, using the near horizon symmetries. Both of these
results, just coming from careful consideration of the near horizon symmetries and have
nothing to do with the regularity of the solution inside the horizon. 
Further more, we find that by varying the non-linear coupling, the regular solution can be
smoothly connected to the 
extremal Reisner-Nordstrom solution of Einstein-Maxwell theory.

The paper is organised as follows. In section~\ref{rbh} we review a particular regular
black hole solution of interest. Then, in section~\ref{sec:ent}, we review the entropy
function formalism and apply it to Einstein gravity coupled to non-linear
electrodynamics. In section~\ref{sec:rbh:E} we consider the special case of the formalism
applied to a regular black hole solution. Finally we end with the conclusion in
section~\ref{sec:conc}, having relegated some technical details about the large charge,
small coupling expansion of the entropy to appendix~\ref{sec:a1}.

\section{Regular Black holes}
\label{rbh}

In this section we review a magnetically charged regular black hole solution of Einstein gravity coupled to
non-linear electrodynamics and its extremal limit
\cite{hep-th/9911174,gr-qc/0006014,gr-qc/0403109,hep-th/0606185}, mainly following
\cite{gr-qc/0403109} with slightly different notation.

 We consider an  action  given by,
\begin{equation}
  \label{Lag}
  S= \frac{1}{16\pi} \int d^4x \sqrt{-g}({R} - {\mathcal L}_{F}(F^2))\; ,
\end{equation}
where $F^2= F_{\mu\nu}F^{\mu\nu}$ and the non-linear $U(1)$ gauge field Lagrangian,
${\mathcal L}_{F}$,
is,\footnote{The coupling $a=\sqrt{\lambda}$ is commonly used in the literature. With out
  loss of generality, we can take $\lambda>0$.}
\begin{equation}
  \label{lm}
  {\mathcal L}_{F} = F^2 \cosh^{-2}\left((\lambda^{2}F^2/2)^{1/4} \right) \;.
\end{equation}
 
The equations of motion corresponding to the metric and gauge field and the Bianchi
identity are,
\begin{eqnarray}
  && R_{\mu\nu} -\frac{1}{2} g_{\mu\nu} R 
  = \frac{\partial {\mathcal L}_{F}}{\partial (F^{2})}2 F_{\mu\lambda}F_{\nu}^{\phantom{\nu}\lambda} 
  -\frac{1}{2} {\mathcal L}_{F} g_{\mu\nu} \; , \\
  && \partial_{\mu} \left(\sqrt{-g} \frac{\partial {\mathcal L}_{F}}{\partial (F^{2})} F^{\mu\nu}\right) =0 \; , \\
  && \partial_{[\mu}F_{\alpha\beta]} =0 \;.
\end{eqnarray}
For a magnetically charged black hole, the equation of motion for the gauge field and the
Bianchi identity can be solved  by,
\begin{equation}
\label{MF}
  F_{\theta\phi} = P \sin\theta,
\end{equation}
where $P$ is the magnetic charge of the black hole. A static, spherically symmetric ansatz for the metric:
\begin{equation}
  ds^2= -a^{2}(r) dt^2 + \frac{dr^2}{a^{2}(r)} + r^2 d\Omega_2^2\; ,
\end{equation}
can solve  Einstein equations with,
\begin{equation}
  a^{2}(r)  = 1-\frac{2m(r)}{r} \; ,
\end{equation}
where,
\begin{equation}
  m(r)= m_\infty- \frac{|P|}{2|\lambda/P|^{1/2}} \tanh \frac{ |\lambda/P|^{1/2}}{r/|P|}.
\end{equation}
The  parameter, $m_\infty$, is an integration constant which can be fixed by employing the boundary
condition $m(\infty)=M$, where $M$ is the black hole mass. Moreover demanding of the
regularity of the line element as $r\rightarrow 0 $, yields,
\begin{equation}
  \label{reg:cond}
  M=\frac{|P|}{2|\lambda/P|^{1/2}},
\end{equation}
and
consequently, $m(r)$ reads,
\begin{equation}
  m(r)=M\left(1-\tanh \frac{P^2}{2Mr}\right) 
  = \frac{|P|}{2|\lambda/P|^{1/2}}\left(1- \tanh \frac{ |\lambda/P|^{1/2}}{r/|P|}\right) \;.
\end{equation}
The location of the inner and outer horizons, $r_{\pm}$,  which are given by equation $a(r)=0$, can be expressed in terms
of the real branches of the Lambert function, $W_i(x)$, as follows,
\begin{equation}
  \frac{r_+}{M} = -\frac{p^2}{W_0(-e^{p^2/4} p^2/4 )-p^2/4} \; , 
  \quad  \frac{r_-}{M} = -\frac{p^2}{W_{-1}(-e^{p^2/4} p^2/4 )-p^2/4}\;,
\end{equation}
where, $p = P/M$, is the magnetic charge-to-mass ratio. 
 The Lambert function\footnote{See \cite{lamb} for a nice review of the properties of the
   Lambert function.}, is defined by the
formula,
\begin{equation}
  \label{lambf}
  e^{W(x)}W(x)=x \;.
\end{equation}
This function has two real branches, called $W_0$ and $W_{-1}$, with the branch point at
$x=-1/e$. Since the value of the principal branch of the Lambert function, $W_{0}$, at $1/e$, plays an
important role in our discussion, we define $w_0 = W_{0}(1/e)$.
 
When $p =p_{ext} = 2 w_0^{1/2}$, $r_{+}=r_{-}$, and the two horizons merge into a
degenerate horizon giving an extremal solution.  Since we will be considering the near
horizon geometry, we will eliminate the mass from our formulae as it is defined asymptotically.
Using (\ref{reg:cond}) we can express the condition for extremality and
regularity, $p_{ext} = 2 w_0^{1/2}$, as,
\begin{equation}
  \label{eq:reg:cond2}
  \frac{\lambda}{P} = w_{0}.
\end{equation}
In other words for an extremal black hole to be regular we must tune the charge to
coupling ratio to a particular value. A generic  extremal, but not necessarily regular,
solution to (\ref{Lag}) will still have a degenerate horizon but presumably with a different charge to coupling ratio. 

One can write the near horizon limit, found by  \cite{gr-qc/0403109} as:
\begin{equation}
\label{nh1}
  ds^2= v_1\left(-\rho^2 dt^2 + \frac{d\rho^2}{\rho^2}\right) + v_2 \left(d\theta^2 + \sin^2\theta d\phi^2\right)\;.
\end{equation}
with,
\begin{align}
  \label{result:rbh}
  v_{2}&= \frac{4w_{0}}{(1+w_{0})^{2}}P^{2}\approx 0.68 P^{2}\;,\\
  v_{1}&= \frac{8w_{0}}{(1+w_{0})^{3}}P^{2}\approx 1.07 P^{2}\;,\\
  \frac{v_{2}}{v_{1}}&=\frac{1}{2}(1+w_{0})\approx 0.64\;,
\end{align}
and the Bekenstein-Hawking entropy is,
\begin{equation}
  S_{BH}=\tfrac{1}{4}A=\pi v_{2}=\frac{4\pi w_{0}}{(1+w_{0})^{2}}P^{2}\;.
\end{equation}

\section{Entropy function Analysis}
\label{sec:ent}

In this section we briefly review the entropy function formalism of Sen
\cite{hep-th/0506177,hep-th/0508042} and subsequently apply it to magnetically charged
extremal solutions of (\ref{Lag}).

Assuming a gauge and diffeomorphism invariant Lagrangian and a near horizon
$AdS_{2}\times S^{2}$ geometry, the entropy function is defined as the Legendre transform,
with respect to the electric charges, of the reduced Lagrangian evaluated at the horizon:
\begin{eqnarray}
\label{EF}
{\mathcal E}({\vec u},{\vec v},{\vec Q}, {\vec P})
= 2\pi\bigg(e^{i}Q_{i}-f({\vec u},{\vec v}, {\vec e},{\vec P})\bigg)
= 2\pi\bigg(e^{i}Q_{i}-\int_{H}
d\theta d\varphi \sqrt{-G}{\cal L}\bigg), 
\end{eqnarray}
where ${e^{i}}$ are the electric fields, $Q_{i}=\partial f/\partial e^{i}$, are the
electric charges conjugate to the electric field, $\vec u$ are the values of the scalar
moduli at the horizon, and $v_1$, $v_2$ are the sizes of the $AdS_2$ and $S^2$,
respectively.   The near horizon  equations of motion
for a black hole carrying electric charges $\vec Q$ and magnetic charges $\vec P$,
are equivalent to the extremisation of ${\cal E}$ with respect to $\vec u,\vec v$ and
$\vec e$:
\begin{eqnarray} \frac{\partial {\mathcal E}}{\partial \vec u}=0\,, \qquad
\frac{\partial {\mathcal E}}{\partial v_i}=0\,, \qquad
\frac{\partial {\mathcal E}}{\partial \vec e}=0\,.
\label{attractor}
\end{eqnarray}

Furthermore, the Wald entropy associated with the black hole is given by ${\mathcal E}$ at
the extremum (\ref{attractor}). If ${\mathcal E}$ has no flat directions, then the
extremization of ${\mathcal E}$ determines ${\vec u}$, ${v_i}$, and ${\vec e}$, in terms
of ${Q}$ and ${P}$. The extremal value of the Wald entropy,
$S={\mathcal E}(\vec Q,\vec P)|_{extr}$, is independent of the asymptotic values of the
scalar fields. This neatly demonstrates the attractor mechanism,
\cite{Ferrara:1995ih,Strominger:1996kf, Ferrara:1996dd}, with out requiring
supersymmetry \cite{9702103}.  The formalism can even be extended to rotating black holes which have less
near horizon symmetry \cite{0606244}. However, since it only involves the near horizon
geometry, a weakness of the formalism is that one implicitly assumes that the full
solution exists, which is not always the case \cite{0507096}.

We now specialise our discussion to the case of interest.  Since the regular black hole
solution has an extremal limit, one can use the entropy function formalism to find the
near horizon geometry and the entropy. We take the near horizon $AdS_{2}\times S^{2}$
metric to be give by (\ref{nh1}).
From the definition (\ref{EF}),  using the Lagrangian (\ref{Lag}), the entropy function
evaluates to,
\begin{equation}
  \label{e}
  {\cal E} = \pi 
  \left(
    v_{2}-v_{1}+\tfrac{1}{2}v_{1}v_{2}{\cal L}_{F}(2P^{2}/v_{2}^{2})
  \right)\; .
\end{equation}
By extremizing this entropy function with respect to $v_1$ and $v_2$, we find following
equations,
\begin{align}
  \label{eom1:0}
  0&=-1+\frac{1}{2}v_{2}{\cal L}_{F}(2P^{2}/v_{2}^{2})\;,\\
  0&=1 +\frac{1}{2}v_{1}\frac{\p}{\p v_{2}}\left[v_{2}{\cal L}_{F}(2P^{2}/v_{2}^{2})\right]\;.
  \label{eom2:0}
\end{align}
Substituting (\ref{eom1:0}) into (\ref{e}) gives,
\begin{equation}
  \label{eq:bh}
  {\cal E}=\pi v_{2}=\tfrac{1}{4}A\; ,
\end{equation}
which is just the Bekenstein-Hawking entropy. This result, which is independent 
of the form of ${\cal L}_{F}$, is to be expected, since, in the
absence of higher derivative terms, the Bekenstein-Hawking and Wald entropies coincide. 

Now, the
equations of motion allow us to determine $v_{1}$ and $v_{2}$ in terms of $P$ and the
coupling $\lambda$. The first equation, (\ref{eom1:0}), determines $v_{2}$, and consequently the entropy, in terms of 
$P$ (and $\lambda$).  Having found $v_{2}$, (\ref{eom2:0}) allows us to determine $v_{1}$ in terms of
$v_{2}$. Consequently, we see that extremising the entropy function completely determines
the entropy and near horizon geometry in terms of  $P$ (and $\lambda$).

We now consider explicitly  finding $v_{1}$ and $v_{2}$ for a particular Lagrangian. 
Using the Lagrangian (\ref{lm}), (\ref{eom1:0}) and (\ref{eom2:0}) become,
\begin{align}
  \label{eom1}
  0&=-1+(P^{2}/v_{2})\cosh^{-2}(\sqrt{\lambda P/v_{2}})\;,\\
  0&=1 -v_{1}\left({P/v_{2}}\right)^{2}\cosh^{-2}(\sqrt{\lambda P/v_{2}})
  \nonumber\\
  &+ v_{1}\sqrt{\lambda}\left({P/v_{2}}\right)^{5/2}\cosh^{-3}(\sqrt{\lambda P/v_{2}})
  \sinh(\sqrt{\lambda P/v_{2}})\;,
  \label{eom2}
\end{align} 
which agrees with the near horizon equations of motion found directly in \cite{gr-qc/0403109}.

\begin{figure}[hbtp] 
 \centering
  \includegraphics[width=0.70\textwidth]{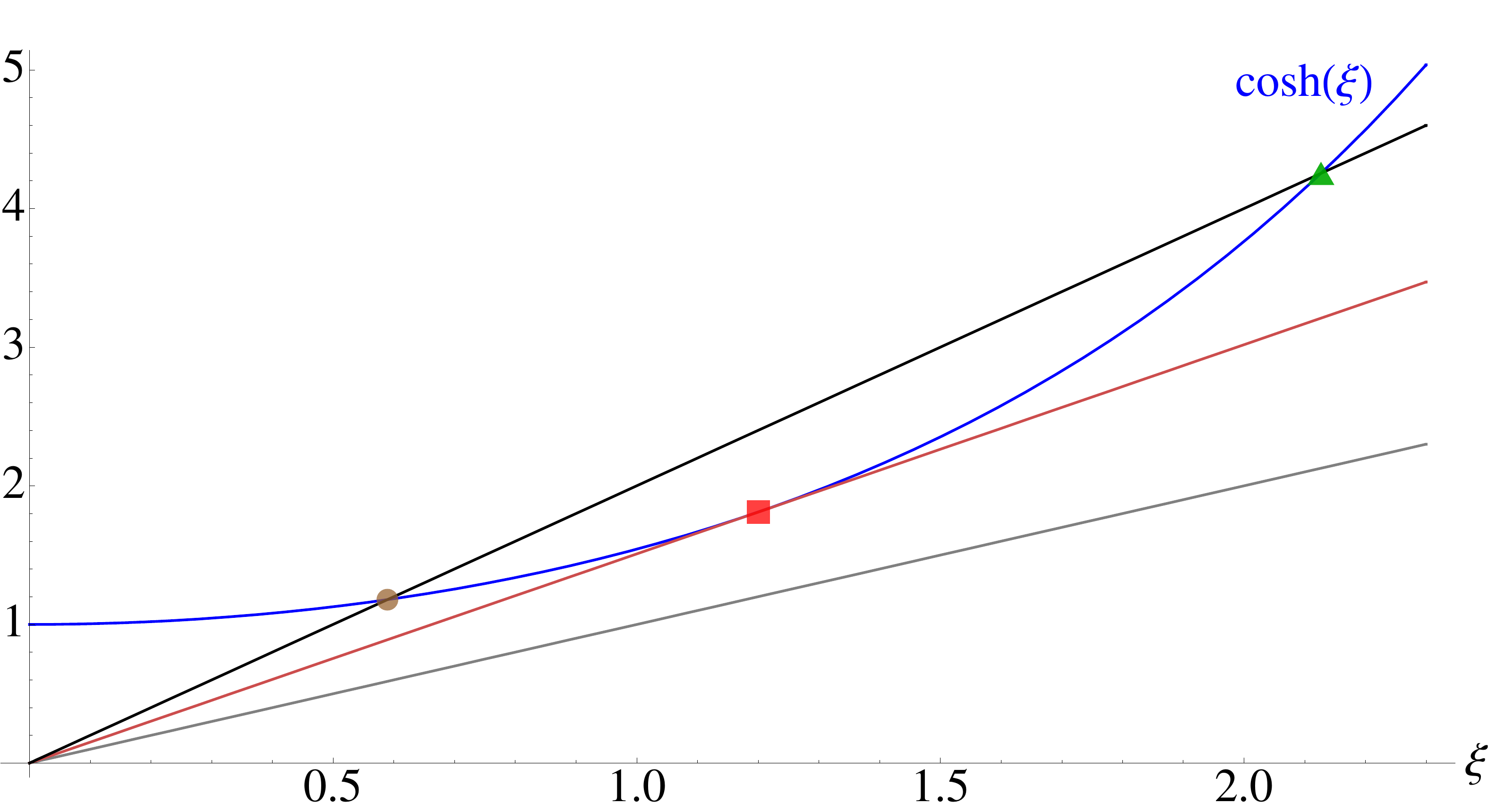}
  \caption{\small This figure illustrates the graphical solution of (\ref{eq:x}) which is given
    by the intersection of $\cosh\xi$ and $\gamma\xi$.  As we increase the gradient,
    $\gamma=\sqrt{P/\lambda}$, one obtains either no solutions, a tangential point
    (denoted by a red square above) or two solutions.  The first point of a double
    intersection, labelled with a brown dot, corresponds to a point on what we call, for
    reasons that will be clear later, the
    large branch and the second intersection, labelled with a green triangle, is on the small
    branch.}
  \label{fig:1}
\end{figure}

To solve (\ref{eom1}), it is
convenient to rewrite it as,
\begin{equation}
  \label{eq:x}
  \cosh\xi = \gamma\xi,
\end{equation} 
where,
\begin{equation}
  \label{eq:def:u:gamma}
 \xi  = \sqrt{\lambda P/v_{2}},\qquad \gamma  = (\lambda/P)^{-1/2}\;.
\end{equation}
One can then graphically solve (\ref{eom1}) by finding the intersection of $\cosh\xi$ and
$\gamma\xi$ for various values of $\gamma$.  We illustrate this procedure in
figure~\ref{fig:1}.  It is not hard to see that as we increase $\gamma$, there are either
zero, one or two solutions to (\ref{eq:x}).  One can also see from figure~\ref{fig:1},
that, as  $\lambda/P\rightarrow0$ (i.e. $\gamma\rightarrow\infty$) , the two possible values for $\xi$ are,
\begin{equation}
  \label{eq:branches}
  {\xi}|_{\frac{\lambda}{P}\rightarrow 0}\rightarrow
\left\{ 
  \begin{array}{c}
    \infty \\ 0
  \end{array}
\right..
\end{equation}
Notice that, since $\cosh x \geq 1 $, (\ref{eom1}) also implies, \begin{equation}
  \label{eq:ineq}
  v_{2}\leq P^{2}.
\end{equation}

\begin{figure}[hbtp]
  \centering
  \includegraphics[width=0.70\textwidth]{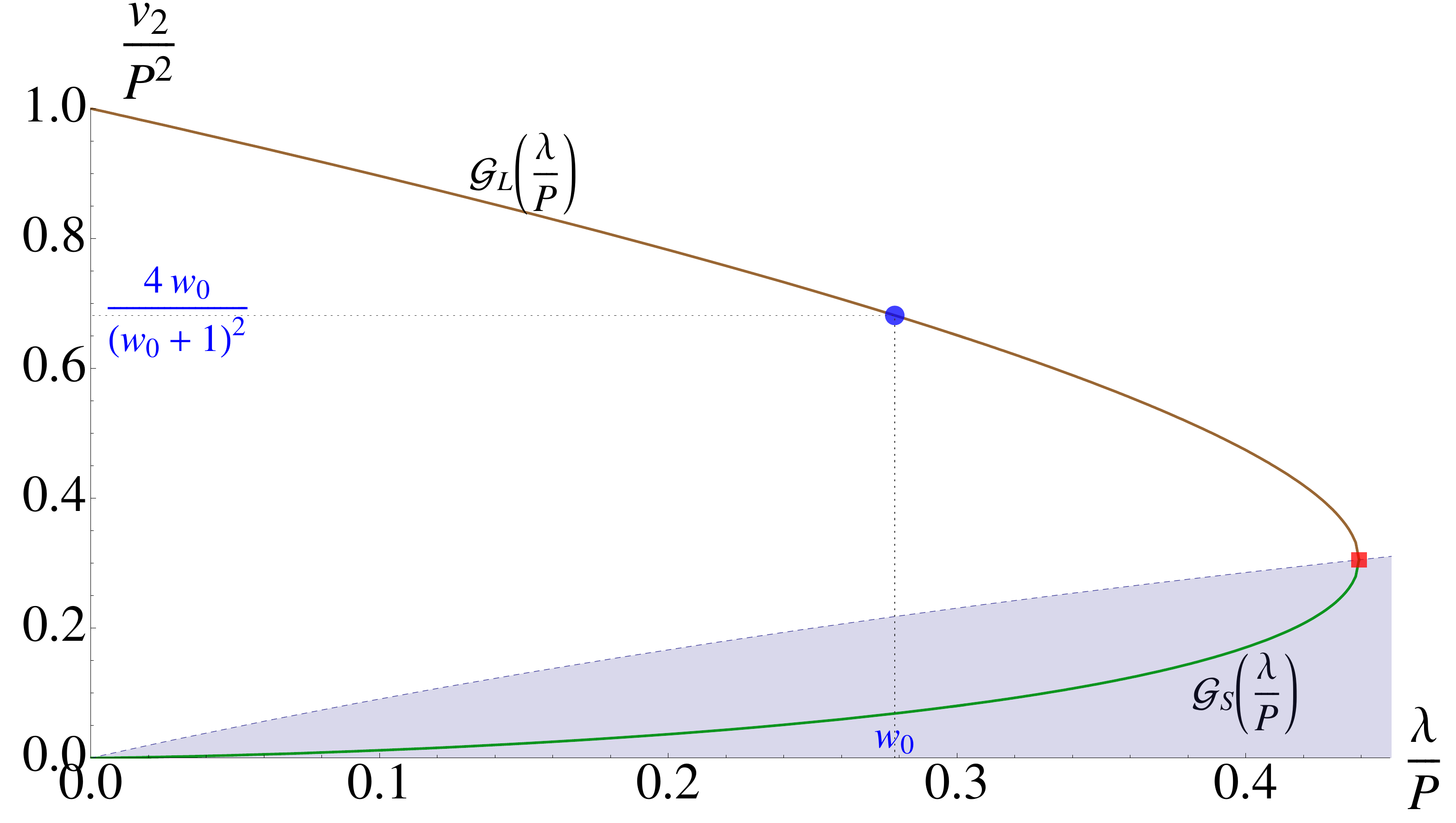}
  \caption{\small This figure shows $v_{2}$ as a function of $\lambda$ and $P$ found by
    numerically solving (\ref{eq:x}). Specifically we plot,
    $v_{2}P^{-2}=\xi^{-2}\gamma^{-2}={\mathcal G}(\lambda/P)$. The large branch,
    ${\cal G}_{L}$, is plotted in brown and the small branch, ${\cal G}_{S}$, is plotted
    in green. We should exclude the shaded region, defined by
    $\gamma{^2}\xi^{2}-\gamma^{2}<1$ , in which, using (\ref{eq:sol:v1b}), $v_{1}$ is
    negative.  Since it is entirely contained within the shaded region, the small branch
    is unphysical. The regular black hole, denoted by a blue dot, is found on the big
    branch at $\lambda/P=w_{0}$. At the place where the branches meet, denoted by a
    red square, $v_{1}\rightarrow\infty$ (or $-\infty$ if we approach from below). }
    \label{fig:2}
\end{figure}

Now, we can define a (multi-valued) function, $\F(x)$, by,
\begin{equation}
  \label{eq:def:F}
  \frac{{\cal F}(x)}{\cosh{{\cal F}(x)}}=\sqrt{x}\;,
\end{equation}
 so that we can formally write down a solution  to (\ref{eq:x}) as,
 \begin{equation}
   \label{eq:sol:u}
  \xi={\cal F}(\gamma^{-2})=\F(\lambda/P)\;.
 \end{equation}
Then letting,
 \begin{align}
   {\cal G}(x)=\frac{x}{\F^{2}(x)} \label{eq:def:ga}
   \refeq{eq:def:F}\frac{1}{\cosh^{2}(\F(x))}\;,
 \end{align}
and using  and (\ref{eq:def:u:gamma}), we can  write,
\begin{equation}
  \label{eq:def:g}
  v_{2}=\frac{\left(\frac{\lambda}{P}\right)}{\xi^{2}}P^{2}= {\cal G}\left(\frac{\lambda}{P}\right)P^{2}.
\end{equation}
which is of the generic form expected by dimensional analysis.
Since (\ref{eq:x}) may have two solutions, ${\cal F}$ and  ${\cal G}$ both have
two branches. Substituting (\ref{eq:branches}) into (\ref{eq:def:ga}) we find that,
in the limit that the non-linear coupling goes to zero (or the charge becomes very large),
\begin{equation}
  \label{eq:g:branches}
  {\cal G}(0)=1/\cosh^{2}({\cal F}(0))=\left\{ 
  \begin{array}{ll}
    0 & \mbox{(small branch, $\G_{S}$)}\\ 1& \mbox{(large branch, $\G_{L}$)}
  \end{array}
\right..
\end{equation}
We call the two branches of ${\cal G}$, the small and large branch.  While it seems
challenging to find an analytical expression for $\G$, it is very easy to evaluate it
numerically. We have plotted $\G$, or in other words $v_{2}/P^{2}$, as a function of
$\lambda/P$ in figure~\ref{fig:2}. We note that $\G$ decreases monotonically on the large
branch, so that, for a fixed charge, the $\lambda=0$ solution is the most entropic.

Having  determined $v_{2}$ (at least in principle),  we can find $v_{1}$ by substituting (\ref{eom1}) into
(\ref{eom2}) and using (\ref{eq:def:g}),  we get,
\begin{align}
  \label{eq:sol:v1}
  v_{1} &= v_{2}(1-\gamma^{-1}\sqrt{\xi^{2}\gamma^{2}-1})^{-1}\\
\label{eq:sol:v1b}
    &= v_{2}(1-[\lambda/P]^{1/2}\sqrt{{\cal G}^{-1}(\lambda/P)-1})^{-1}\;,
\end{align}
with (\ref{eq:ineq}) ensuring reality.

\begin{figure}[hbtp]
  \centering  
  \includegraphics[width=0.75\textwidth]{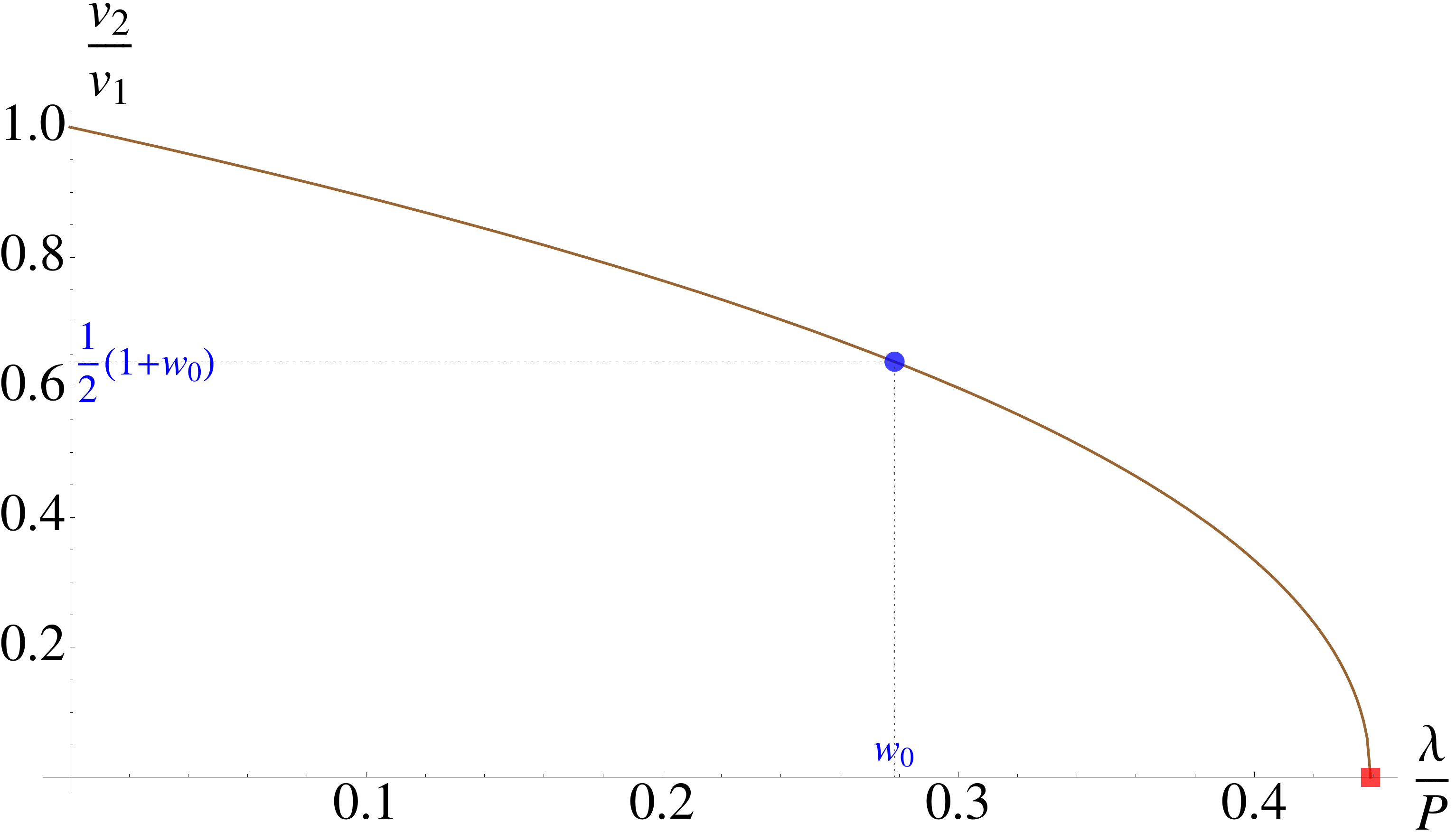}
  \caption{\small This figure shows $v_{2}/v_{1}$ as a function of $\lambda/P$ found by
    numerically solving (\ref{eq:x}) and using (\ref{eq:sol:v1}).  For $\lambda/P$ small,
    we have $v_{1}\approx v_{2}$. As we approach the branch point, denoted by a red
    square, $v_{1}$ diverges and $v_{2}/v_{1}\rightarrow0$. The regular black hole, at
    $\lambda/P=w_{0}$, is denoted by a blue dot.  }
    \label{fig:3}
\end{figure}

For $v_{1}$ to be positive and finite, on sees that from (\ref{eq:sol:v1}), we require
$\xi^{2}\gamma^{2}-1> \gamma^{2}$.  Now, at the branch point, the function
$f(\xi)=\cosh\xi - \gamma\xi$ has a single zero, so we require that $f'(\xi)=0$ when
$f(\xi)=0$. In other words, in addition to (\ref{eq:x}) the branch point is determined by,
\begin{equation}
  \label{eq:fdash}
  \sinh\xi = \gamma.
\end{equation}
Combining (\ref{eq:x})and (\ref{eq:fdash}) gives,
\begin{equation}
  \gamma^{2}\xi^{2}-1=\gamma^{2}\;,
\end{equation}
which, using (\ref{eq:sol:v1}), implies that as we approach the branch point, $v_{1}\rightarrow\infty$ or in other
words the $AdS_{2}$ approaches flat space. From figure~\ref{fig:2}, we see that the small
branch lies entirely in the region $\gamma{^2}\xi^{2}-\gamma^{2}<1$, and consequently,
$v_{1}$ is always negative on it, making it unphysical. Discarding the small branch, we
have used~(\ref{eq:sol:v1b}) to plot $v_{2}/v_{1}$ as a function of $\lambda/P$ in
figure~\ref{fig:3}. We see that $v_{1}/v_{2}$ increases monotonically, eventually diverging
at the branch point.

Finally, one can actually obtain a large charge, small coupling expansion for
$\G_{L}$. Assuming $\G_{L}$, has a nice Taylor expansion about zero, using
(\ref{eq:g:branches}) as a starting point, and by taking successive derivatives of
(\ref{eq:def:F}) and (\ref{eq:def:ga}) one can recursively expand $\G_{L}(x)$ about
zero. As discussed in appendix~\ref{sec:a1}, we find that,
\begin{align}
\G_{L}(x)&= 1-x-\tfrac{1}{3}x^{2}
+{\cal O}\left(x^{3}\right)\; ,
 \label{taylor:L}
\end{align}
so that we can write a large charge, small coupling expansion for the entropy,
\begin{align}
{\cal E}= \pi P^{2}\left( 1-\frac{\lambda}{P}-\frac{1}{3}\frac{\lambda^{2}}{P^{2}}
+{\cal O}\left(\frac{\lambda^{3}}{P^{3}}\right)\right) \;.
 \label{largeP}
\end{align}
For completeness, we mention that on the small branch, as discussed in appendix~\ref{sec:a1}, for $x$
small, we get,
\begin{equation}
  \label{eq:gs:approx}
 \G_{S}(x)\approx x{\left[ W_{-1}(-\tfrac{1}{2}\sqrt{x})\right]^{-2}}\;.
\end{equation}
where $W_{-1}$ is the non-principal real branch of the Lambert function.

As a check we note that on the large branch, taking $\lambda\rightarrow0$,
we recover the usual near horizon extremal Einstein-Maxwell
Reisner-Nordstrom solution with,
\begin{align}
  \label{result}
  v_{1}=v_{2}= P^{2}\;.
\end{align}

\section{Entropy function and the regular black hole}
\label{sec:rbh:E}

In this section we confirm that  the entropy function analysis of the regular black hole
reproduces the near horizon geometry of the known solution found in \cite{gr-qc/0403109}.
This merely entails considering the results of the previous section with the appropriate
value of $\lambda/P$. 

As discussed in section~\ref{rbh}, the regular black hole corresponds to the point
$\lambda/P={w_{0}}$, so that (\ref{eq:x}) becomes,
\begin{equation}
  \label{eq:rbh:eom1}
  \cosh\xi = {w_{0}}^{-1/2}\xi. 
\end{equation}
One can analytically check, using the property $w_{0}e^{w_{0}}=e^{-1}$, that (\ref{eq:rbh:eom1}) has a
solution, 
\begin{equation}
  \label{u:rbh}
 \xi=\frac{w_{0}+1}{2},
\end{equation}
or in other words $\F(w_{0})=\frac{1}{2}(w_{0}+1)$. We have plotted the position of the
regular  solution as a blue dot in figures~\ref{fig:2} and~\ref{fig:3}, from which we
observe that it is on the large branch.

Finally, one can check that substituting the solution, (\ref{u:rbh}), into (\ref{eq:def:g}) and
(\ref{eq:sol:v1}) reproduces (\ref{result:rbh}) and we are done.

\section{Conclusion}\label{sec:conc}

In this paper we examine entropy function formalism for regular magnetically charged black
hole solution in Einstein-Hilbert gravity coupled with a certain non-linear $U(1)$ gauge
field. The mass and charge of the full solution can be tuned so that it has no curvature
singularity at the centre. In the extremal limit this corresponds to a particular charge
to non-linear coupling ratio with an $AdS_2\times S^2$ near horizon
geometry. Unsurprisingly we find that the entropy function analysis match with the exact
solution found by solving the full Einstein equations.  This is in contrast with the claim
in the recent papers \cite{0705.2478,0707.1933}.

Indeed in the entropy function formalism, the equation of motion, which follow from
extremizing the entropy function, are exactly the same as equation of motion at horizon
found by extremizing the action, simply because the entropy function (up to Legendre
transformation) is the Lagrangian at the horizon. The fact that the entropy is the value
of entropy function at its extremum is derived from Wald entropy formula, using the near
horizon symmetries. Both of these results apparently have nothing to do with the regularity of the
solution inside the horizon.

\bigskip {\bf Acknowledgements:} We would like to thank Ashoke Sen for helpful
comments. The work of K.G. is, in part, supported by the EU-RTN network contract
MRTN-CT-2004-005104 and INTAS contract 03-51-6346. The work of H.Y is supported by the
Korea Research Foundation Leading Scientist Grant (R02-2004-000-10150-0) and Star Faculty
Grant (KRF-2005-084-C00003).

\appendix

\section{Large charge/small coupling expansion of  the entropy}
\label{sec:a1}

In this appendix we discuss the expansion of $\G(x)$ about zero.

Taking the derivative of (\ref{eq:def:F}) with respect to $x$ and solving for $\F'$ we
find,
\begin{equation}
  \label{eq:a1}
  \F'(x)=\frac{\cosh (\F(x))}{2 \sqrt{x} \left(\sqrt{x} \sinh (\F(x))-1\right)}\;,
\end{equation}
while taking of derivative of (\ref{eq:def:ga}) gives,
\begin{equation}
  \label{eq:a2}
  \G'(x)=-2 \text{sech}^2(\F(x)) \tanh (\F(x)) \F'(x)\;.
\end{equation}
Now using (\ref{eq:def:F}) and (\ref{eq:a1}) we can rewrite (\ref{eq:a2}) as,
\begin{equation}
  \label{eq:a3}
  \G'(x)= \frac{\tanh (\mathcal{F})}{\mathcal{F} (\mathcal{F} \tanh (\mathcal{F})-1)}\;,
\end{equation}
and using (\ref{eq:branches}) and taking the limit $x\rightarrow0$, we get,
\begin{equation}
  \label{eq:a4}
  {\cal G}'(0)=\left\{ 
  \begin{array}{ll}
    0 & \mbox{(small branch)}\\ -1& \mbox{(large branch)}
  \end{array}
\right..
\end{equation}
Taking another set of derivative, after some algebra we obtain,
\begin{equation}
  \label{eq:a5}
       \G''(x)
       = \frac{\mathcal{F}\cosh (2 \mathcal{F})-\cosh (\mathcal{F}) \sinh (\mathcal{F})}
              {2 \mathcal{F}^3 (\mathcal{F} \tanh (\mathcal{F})-1)^3}\;,
\end{equation}
and once again taking the $x\rightarrow0$ limit, we get,
\begin{equation}
  \label{eq:a6}
  \G''(0)=
  \left\{ 
  \begin{array}{ll}
    \infty & \mbox{(small branch)}\\ -\tfrac{2}{3}& \mbox{(large branch)}
  \end{array}
\right. \;.
\end{equation}
We found that the second order Taylor expansion of $\G_{L}(x)$, (\ref{taylor:L}), agrees well
with our numerical plot for $x$ small. For example, at $x=w_{0}$, we find that they differ
by about $2\%$. 

On the other hand, we see that the small branch does not have a nice Taylor expansion
about the origin. However, on the small branch, we see from figure~\ref{fig:1}, that when $\xi$ is
large, $\gamma$ is also large and consequently, $\lambda/P=\gamma^{-2}$, is small. For $\xi\gg 1$, we can
approximate (\ref{eq:x}) by,
\begin{equation}
  \label{eq:a7}
  \tfrac{1}{2}e^{\xi}\approx\gamma\xi + {\cal O}(e^{-\xi})\;,
\end{equation}
which can approximately be solved  by,
\begin{equation}
 \xi \approx - W(-\tfrac{1}{2}\gamma^{-1})\;.
\end{equation}
where $W$ is the Lambert function defined in (\ref{lambf}).
For, $-e^{-1}<x<0$, the two real branches of $W$ satisfy $W_{0}(x)\geq -1$ and
$W_{-1}(x)\leq-1$, \cite{lamb}, consequently
since we are assuming that $\xi\gg 1$, we should take the branch $W_{-1}$. So, using
(\ref{eq:sol:u},\ref{eq:def:ga}), we obtain,
\begin{equation}
  \G_{S}(\lambda/P)\approx (\lambda/P){\left[ W_{-1}(-\tfrac{1}{2}\sqrt{\lambda/P})\right]^{-2}}\;,
\end{equation}
which we found agrees well with our numerical results shown in figure~\ref{fig:2},  for $\lambda/P$ small.

\bibliographystyle{JHEP}\bibliography{nl4}

\end{document}